\documentclass[%
 reprint,
 amsmath,amssymb,
 aps,
]{revtex4-1}

\usepackage{graphicx}
\usepackage{dcolumn}
\usepackage{bm}
\usepackage{hyperref}
\hypersetup{colorlinks=true,linkcolor=blue,filecolor=blue,urlcolor=blue}

\usepackage[
margin=0.5in,
]{geometry}

\begin{document}

\preprint{APS/123-QED}

\title{Properties  of normal modes in a modified disordered Klein-Gordon lattice: \\From disorder to order}

\author{B.~Senyange}
\author{J.-J.~du Plessis}
\author{B.~Many Manda}
\author{Ch.~Skokos}%
\email{haris.skokos@uct.ac.za}
\affiliation{Department of Mathematics and Applied Mathematics,
	University of Cape Town, Rondebosch, 7701 Cape Town, South Africa.}

\date{\today}

\begin{abstract}
We introduce a modified version of the disordered  Klein-Gordon lattice model, having two parameters for controlling the disorder strength: $D$, which determines the range of the coefficients of the on-site potentials, and $W$, which defines the strength of the nearest-neighbor interactions. We fix $W=4$ and investigate how the properties of the system's normal modes change as we approach its ordered version, i.e.~$D\rightarrow 0$. We show that the probability density distribution of the normal modes' frequencies takes a `U'-shaped profile as $D$ decreases. Furthermore, we use two quantities for estimating the modes' spatial extent, the so-called localization volume $V$ (which is related to the mode's second moment) and the mode's participation number $P$. We show that both quantities scale as $\propto D^{-2}$ when $D$ approaches zero and we numerically verify a proportionality relation between them as $V/P \approx 2.6$.
\end{abstract}

\maketitle


\section{Introduction}
\label{sec:intro}

The phenomenon where energy propagation is halted in linear disordered media
was discovered over $50$ years ago by Anderson \cite{Anderson1958} and is referred to as Anderson Localization (AL). In recent years AL has been extensively investigated in experimental
\cite{Hu2008,Lahini2008,Cobus2016}, numerical and theoretical \cite{Kopidakis2008,Pikovsky2008,Flach2009,Laptyeva2014} studies.
Experiments on the observation of AL include light propagation in spatially random optical media \cite{Schwartz2007,Lahini2008},  non-interacting Bose-Einstein condensate expansions in random optical potentials \cite{Billy2008,Raoti2008}, as well as  wave localization in a microwave cavity filled with randomly distributed scatterers \cite{Dalichaouch1991}.

In linear disordered systems with sufficiently strong disorder, all normal modes (NMs) are localized and any wave packet which is
initially localized remains in that state for all time. On the other hand, the introduction
of nonlinearities to such systems leads to the interaction
of the NMs and the introduction of chaos.  In general two typical one-dimensional (1D)
disordered nonlinear lattices, namely the Klein-Gordon model (DKG)
and the discrete
nonlinear Schr\"{o}dinger equation (DDNLS), have been at the center of studies of the effect nonlinearity on AL \cite{Pikovsky2008,Flach2009,Skokos2009,Flach2010,Skokos2013,Laptyeva2014,SMS18}. In these works it was found that eventually nonlinearity destroys AL and the characteristics of different spreading behaviors (the so-called `weak' and `strong chaos' regimes) were identified. The appearance of these spreading regimes depends on the properties of the systems' NMs and in particular the width $\Delta$
of their frequency band, the NMs' localization volume $V$, which quantifies the number of sites where the NM has significant contribution,  the average spacing $d$ of the modes which strongly interact with a particular NM, and the relation of these quantities with the frequency shift $\delta$ induced by the introduction of nonlinearity \cite{Flach2009,LBKSF10,Flach2010,Laptyeva2014}.

In this work we investigate the properties of the NMs of a 1D linear version of the DKG lattice  where two parameters determine the model's disorder strength: $D$, which is related to the on-site potential and $W$, which specifies the interaction between nearest neighbors. In particular, we study  how these properties change when we start from a disordered version of the system and, by changing $D$ (while $W$ is kept constant), we move toward the  ordered system.
We focus our attention on the spatial
structure of the modes and analyze the NMs' frequencies, their localization volume $V$ and participation number $P$ (which provides information about the number of highly excited sites in the NM), along with the changes in the frequency scales $\Delta$ and $d$ of the  system.

The paper is organized as follows. In Sect.~\ref{sec:models} we describe in detail the modified Klein-Gordon system we consider in this study, along with its relation to the well-known linear disordered  Schr\"{o}dinger equation (LDSE). Then, in Sect.~\ref{sec:num_res} we present a detailed numerical study of the properties of the  system's NMs and their changes as  $D$ decreases, emphasizing the distributions of the NMs' frequencies and some measures of their spatial extent. Finally, in Sect.~\ref{sec:summary} we summarize our results and discuss their significance.

\section{The modified Klein-Gordon model}
\label{sec:models}

In this work we perform an analysis of the NMs of the linear disordered Klein-Gordon (LDKG) model whose Hamiltonian function is
\begin{equation}
	H_K = \sum _{l} \left[ \frac{p_l^2}{2} + \epsilon_l \frac{q_l^2}{2} + \frac{1}{2W}\left(q_{l+1} - q_l\right)^2 \right],
	\label{eq:LDKG}
\end{equation}
with $q_l$ and $p_l$ respectively representing the generalized
position and momentum of the $l$-th oscillator in a
chain of $N$ particles.
The coefficients $\epsilon_l$ take uncorrelated random values
chosen from the uniform probability distribution function
$\mathcal{P}(\epsilon_l)=1/(2D)$ in the interval $\left[ 1-D, 1+D\right]$, where $D$ is a parameter defining the width of the disorder range, and $W >0$ determines the strength of the hopping. In particular, we set $0 \leq D \leq 1/2$, and we consider fixed boundary
conditions: $q_1=q_{N+1}=p_1=p_{N+1}=0$. We note that when $D$ tends to zero, Hamiltonian \eqref{eq:LDKG} shifts from  a disordered
system toward an ordered one without affecting the strength ($1/W$) of the interactions between nearest neighbors. Setting $\epsilon_l=1$ for all sites $l=1,\ldots,N$ in \eqref{eq:LDKG} gives an ordered linear model, whose nonlinear version has been studied in
\cite{Danieli2019}.

The  equations of
motion of system (\ref{eq:LDKG}) are
\begin{equation}\label{eq:eqmot_LDKG}
\ddot{q_l}= -\left[\epsilon_lq_l+
\frac{1}{W}\left(2q_l-q_{l-1}-q_{l+1}\right)\right],
\end{equation}
where a dot denotes the derivative with respect to time $t$. By  using the ansatz $q_l=A_l\exp(i\omega t)$,
where $A_l$ is the amplitude of oscillator $l$, Eq.~\eqref{eq:eqmot_LDKG} leads to the eigenvalue problem
\begin{equation}\label{eq:eigval_LDKG}
	\omega^2A_l=\frac{1}{W}\left[(W\epsilon_l+2)A_l-A_{l-1}-A_{l+1}\right].
\end{equation}
The normalized eigenvectors $A_{\nu,l}$, $\nu =1,2, \ldots,N$, with $\sum_lA_{\nu,l}^2 = 1$,
are the system's NMs and the eigenvalues $\omega^2_{\nu}$ are the corresponding
squared frequencies of these modes.

The dynamics of the nonlinear version (DKG) of system \eqref{eq:LDKG} obtained by the presence of an additional nonlinear term in the on-site potential ($\sum_l q_l^4/4$) has been extensively studied \cite{Flach2009,Skokos2009,LBKSF10,Skokos2013,SMS18}, mainly in comparison with the DDNLS model, i.e.~the nonlinear version of the LDSE
\begin{equation}
	\label{eq:LDSE}
	H_D = \sum_l \left[ \tilde{\epsilon}_l \lvert \psi_l \rvert ^2  - \left(\psi_{l + 1}^\star \psi _l + \psi_{l + 1}\psi _l^\star\right) \right],
\end{equation}
where $\psi_l$ is the complex wave function at site $l$, the $(^\star)$ denotes the complex conjugate, and $\tilde{\epsilon}_l$ is a random number drawn uniformly from an interval symmetrically located around zero. By setting this interval to be $\left[ -\mathcal{W}/2, \mathcal{W}/2 \right]$, with $\mathcal{W}\geq 0$ denoting the disorder strength, system \eqref{eq:LDSE} corresponds to the standard tight-binding (i.e.~nearest-neighbor hopping) Anderson model with disorder on the on-site potentials \cite{Anderson1958,Kramer1993}. We note that the DDNLS system studied in \cite{Flach2009,Skokos2009,LBKSF10,Skokos2013,SMS18} is obtained by the addition of the term $\beta \lvert \psi_l \rvert ^4/2$ in \eqref{eq:LDSE}, with $\beta \geq 0$ quantifying the nonlinearity strength. The NMs of system \eqref{eq:LDSE} can be found by setting $\psi_l=A_l\exp(-i\lambda t)$, which leads to the linear eigenvalue problem
\begin{equation}\label{eq:eigval_LDSE}
	\lambda A_l=   \tilde{\epsilon}_l A_l-A_{l-1}-A_{l+1}.
\end{equation}

The eigenvalue problems \eqref{eq:eigval_LDKG} and \eqref{eq:eigval_LDSE}, and consequently the Hamiltonian systems \eqref{eq:LDKG} and \eqref{eq:LDSE}, become identical for
\begin{eqnarray}
	\label{eq:change_lambda}
	\lambda &=& \omega ^2 W - W - 2, \\
	\label{eq:change_epsilon}
	\tilde{\epsilon}_l &=& W\left(\epsilon_l - 1\right),\\
	\label{eq:change_W}
	\mathcal{W} &=& 2DW.
\end{eqnarray}
Solving the eigenvalue problem \eqref{eq:eigval_LDKG} is equivalent to diagonalizing
the $N \times N$ tridiagonal matrix ${\bf A}$ with elements
\begin{equation}
\label{eq:A_epsilon}
	a_{l,l-1}=a_{l,l+1}=-\frac{1}{W},\;\; a_{l,l}=\epsilon_l+\frac{2}{W},
\end{equation}
and $a_{k,l} = 0$ otherwise, for $l, k=1,2, \ldots,N$. Bounds of the eigenvalues $\omega^2$ of ${\bf A}$ can be found by applying the \textit{Gershgorin circle
theorem} \cite{Gerschgorin1931}, which states \cite{Wolkowicz1980} that the
	eigenvalues $\omega^2$ of
matrix ${\bf A}$ are bounded as $\left| \omega^2-a_{l,l} \right| \leq R_l-\left|a_{l,l} \right|$, $l=1,2,\ldots,N$ with $R_l=\sum_{k=1}^{N}\left| a_{l,k} \right|$. The direct application of this theorem to  matrix \eqref{eq:A_epsilon} gives $\epsilon_l \leq \omega^2 \leq \epsilon_l+4/W$. Since
$\epsilon_l\in [1-D,1+D]$, then the minimum ($\omega_{\nu,-}^2$) and
maximum ($\omega_{\nu,+}^2$) values of the squared frequencies are $\omega_{\nu,-}^2=1-D$
and $\omega_{\nu,+}^2=1+D+4/W$. Therefore, the width of the squared eigenfrequency spectrum is
\begin{equation}
\label{eq:Delta_K}
	\Delta _K = 2D + \frac{4}{W}.
\end{equation}
Setting $\epsilon_l=1$, $l=1,2,\ldots, N$ in \eqref{eq:LDKG}, i.e.~$D=0$, we obtain  an ordered linear system, whose eigenvalue problem can be solved analytically \cite{Yueh2005,Borowska2015}  giving
\begin{equation}
\label{eq:ordereigval}
\omega_{\nu}^2=1+\frac{2}{W}\left[1-\cos\left(\frac{\nu \pi}{N+1}\right)\right], \,\,\, \nu = 1, 2, \ldots, N.
\end{equation}
We note that $\omega_{\nu}^2$ belongs to the interval  $\left( 1,1 +4/W\right)$, whose width is  $\Delta _K = \frac{4}{W}$ in accordance to \eqref{eq:Delta_K}.

The introduction of the width $D$ of the disorder range in the LDKG model \eqref{eq:LDKG} as an additional parameter gives us the ability to alter the system's disorder strength in two distinct ways, i.e.~by modifying $D$ and/or $W$, while in the equivalent LDSE system \eqref{eq:LDSE} we have only one parameter, $\mathcal{W}$, to change the disorder strength. In the investigations of the DKG (and the LDKG) model \cite{Flach2009,Skokos2009,LBKSF10,Skokos2013,SMS18} performed to date, $D$ was set to $D=1/2$ and typically values $W\geq 1$ were used, which also correspond through \eqref{eq:change_W} to  $\mathcal{W}\geq 1$. In that set-up, the way to study the transition to a more ordered system is to let $W \rightarrow 0$, as this leads to the huge increase of the significance of the last term of Hamiltonian \eqref{eq:LDKG} (i.e.~the nearest-neighbor interactions) over the on-site potential $\epsilon_l q_l^2/2$ which becomes negligible. The introduction of the parameter $D$ allows us to obtain this transition for the  LDKG system \eqref{eq:LDKG} [and equivalently for the LDSE \eqref{eq:LDSE} one] by altering the on-site potentials through $D \rightarrow 0$, while $W$ can be kept fixed.

The properties of the LDSE's NMs were discussed in \cite{Krimer2010} for  $\mathcal{W} \gtrsim 1$ as that work was mainly focused on strong disorder.  Equations \eqref{eq:change_lambda}--\eqref{eq:change_W} allow the direct translation of results obtained for the LDSE \eqref{eq:LDSE} \cite{Kramer1993,Krimer2010} to the case of the LDKG Hamiltonian \eqref{eq:LDKG}. For instance, the
asymptotic spatial decay of NMs of system \eqref{eq:LDKG} is given
by $A_{\nu,l}\sim \exp(-|l-l_0|/\xi_{\nu})$, where
$l_0$=$\sum_l lA_{\nu,l}^2$ is the NM's mean spatial position and
$\xi_{\nu}$ is the so-called localization length of mode $\nu$
\cite{Anderson1958,Kramer1993,Krimer2010}, which, using \eqref{eq:change_W}, is given by
\begin{equation}
	\xi_{\nu}  = \frac{24 \left[4 - (\omega_{\nu}^2 W - W - 2)^2\right]}{4D^2W^2}.
	\label{eq:loc_length_LDKG}
\end{equation}
It is worth noting that $D$ and $W$ affect differently $\Delta_K$ \eqref{eq:Delta_K} and $\xi_{\nu}$ \eqref{eq:loc_length_LDKG}, as they do not appear in these expressions always as a product $DW$. The NMs  with the largest localization length, i.e.~the most extended ones, appear at the
bandwidth center \cite{Kramer1993,Krimer2010} having
\begin{equation}
	\xi \left(\omega_{\nu}^2 = 1 + \frac{2}{W}\right) = \xi _0 = \frac{24}{D^2W^2},
	\label{eq:loc_length_LDKG_max}
\end{equation}
for $\mathcal{W}\leq4$.

\section{Numerical results}
\label{sec:num_res}

In our analysis we set $W=4$, a typical value used in several studies \cite{Flach2009,Skokos2009,LBKSF10,Skokos2013,SMS18}, and change $D$ from $D=1/2$, which corresponds to the $D$ of the disordered model considered in those papers,
to values very close to $D=0$ (ordered system). For this setup Eq.~\eqref{eq:change_W} gives $\mathcal{W}=8D$, while the width of the squared frequency band \eqref{eq:Delta_K} becomes $\Delta_K=2D+1$.
We obtain numerical results by considering lattices of
lengths $N=10{,}000 - 50{,}000$. For a particular value of $D\in(0,0.5]$, we perform
simulations for $n_d=100$ disorder realizations in order to statistically analyze the NMs' properties. We order the NMs either by increasing value of their mean spatial position
$l_0$, or  of their squared frequency $\omega^2$. As the width of the frequency band and its boundaries change with $D$, we  present results according to the NM's normalized squared frequency
\begin{equation} \label{eq:nomr_freq}
	\omega_{\nu, n}^2 = \frac{\omega_{\nu}^2 - \omega_{\nu, -}^2}{\Delta _K} = \frac{\omega^2_{\nu}+D-1}{2D+1},
\end{equation}
in order to make direct comparisons between cases with different $D$ values.

\begin{figure}
	\includegraphics[width=0.5\textwidth,keepaspectratio]{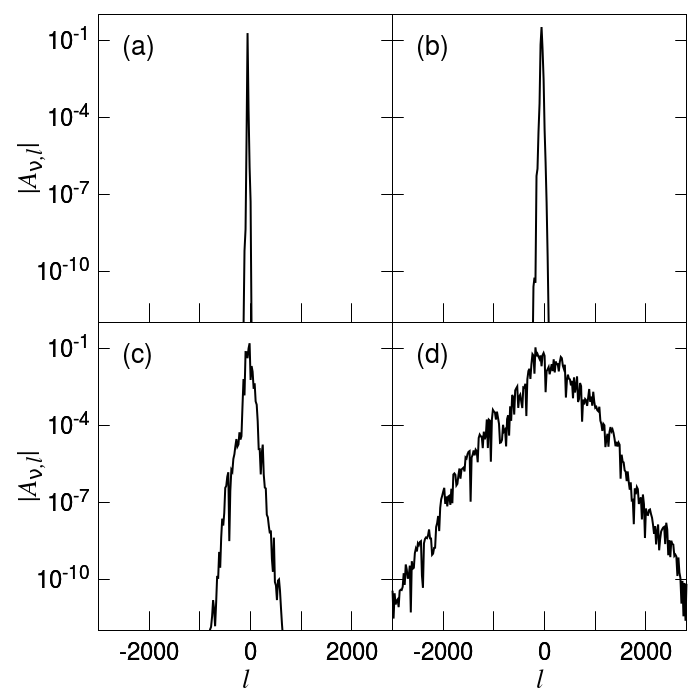}
    	\caption{\label{fig:NMs_profiles} The profile of representative NMs, whose mean spatial position $l_0$ is
		at the center of the lattice and their normalized squared frequencies are $\omega_{\nu, n}^2 \approx 0.5$ for (a) $D=0.5$, (b)
		$D=0.3$, (c)  $D=0.2$  and (d)	$D=0.1$.}
\end{figure}

In Fig.~\ref{fig:NMs_profiles} we present the profiles, i.e.~absolute value of amplitude $A_{\nu,l}$ versus lattice site $l$, of some representative NMs in logarithmic-linear scales for various values of $D$. For all cases the same disordered realization is used, whose values are scaled appropriately to fit the interval $[1-D, 1+D]$. In all panels of Fig.~\ref{fig:NMs_profiles} we plot NMs whose mean spatial position, $l_0$, is close to the lattice center and have approximately the same normalized squared frequency, $\omega _{\nu, n}^2 \approx 0.5$, in the middle of the frequency band where the most extended NMs are \cite{Anderson1958,Kramer1993,Krimer2010}. From the results of Fig.~\ref{fig:NMs_profiles} we see that all NMs are exponentially localized, as they are characterized by clearly defined exponential tails, but their extent is increasing as $D$ decreases, i.e.~as system \eqref{eq:LDKG} becomes less disordered, reaching an extent of a few thousand sites for $D=0.1$ [Fig.~\ref{fig:NMs_profiles}(d)].

\begin{figure}
	\includegraphics[width=0.3\textwidth,keepaspectratio]{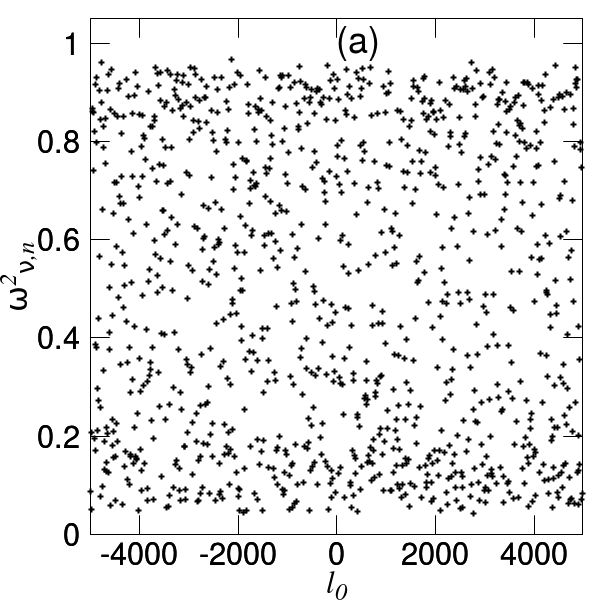}
	\includegraphics[width=0.5\textwidth,keepaspectratio]{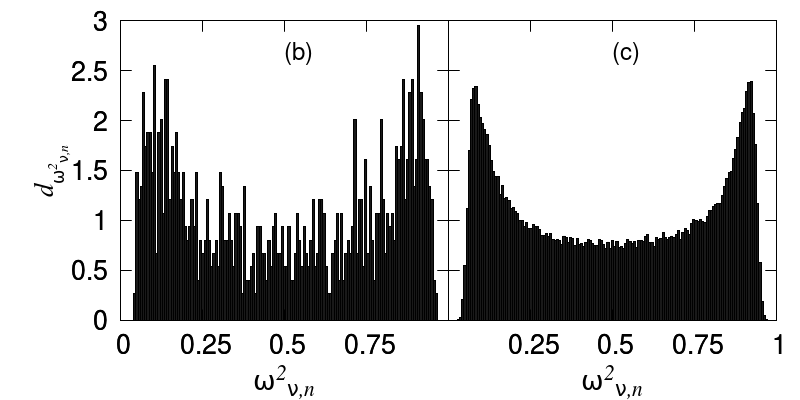}
	\caption{\label{fig:freq_1} (a) The normalized squared frequencies $\omega_{\nu, n}^2$ \eqref{eq:nomr_freq} of the NMs for 1 disorder realization of system \eqref{eq:LDKG} with $N=10{,}000$ and $D=0.1$, as a function of the NMs' mean spatial position $l_0$.  (b) The probability density distribution $d_{\omega _{\nu, n}^2}$ of $\omega_{\nu, n}^2$ of panel (a). (c) Similar to (b) but for $n_d=100$ disorder realizations.}
\end{figure}

In Fig.~\ref{fig:freq_1}(a) we show the normalized squared frequencies $\omega_{\nu, n}^2$ \eqref{eq:nomr_freq} of the NMs for one disorder realization of system \eqref{eq:LDKG} with $N=10{,}000$ and $D=0.1$, as a function of the NMs' mean spatial position $l_0$. We see that throughout the lattice the frequencies are mainly concentrated at the borders of the spectrum as more points are located in the regions $\omega_{\nu, n}^2 \approx 0.1$ and $\approx 0.9$. This feature becomes more evident in Fig.~\ref{fig:freq_1}(b) where we present as a histogram the probability density distribution $d_{\omega _{\nu, n}^2}$ of the frequencies $\omega_{\nu, n}^2$ of Fig.~\ref{fig:freq_1}(a). The maxima at the edges of the distribution are clearly seen. The squared frequency distribution becomes smoother by considering results over $n_d=100$ disorder realizations for $D=0.1$ [Fig.~\ref{fig:freq_1}(c)]. Here the `U' shape of the distribution with equally high peaks at the two edges is seen. From the results of Fig.~\ref{fig:freq_1}(c) we note that the frequencies avoid the extreme ends of the band, i.e.~$\omega_{\nu, n}^2 \approx 0$ and $\approx 1$, something which was also seen in Fig.~\ref{fig:freq_1}(a) where the frequencies of only one disorder realization were presented.

\begin{figure}
	\includegraphics[width=0.4\textwidth,keepaspectratio]{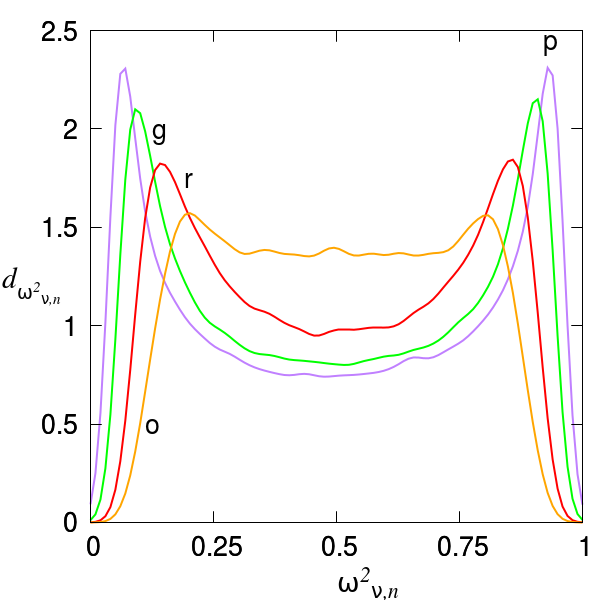}
     \caption{\label{fig:distributions_many}
        The probability density distributions $d_{\omega _{\nu, n}^2}$ of the normalized squared frequencies $\omega_{\nu, n}^2$ \eqref{eq:nomr_freq} of system \eqref{eq:LDKG} for $D=0.5$ (orange - `o'), $D=0.2$ (red - `r'), $D=0.1$ (green - `g') and
		$D=0.06$ (purple - `p'). The results of each curve were obtained from the analysis of $n_d=100$ disorder realizations.}
\end{figure}

In Fig.~\ref{fig:distributions_many} we show how the probability density distribution (over $n_d=100$ disorder realizations) $d_{\omega _{\nu, n}^2}$ of the  $\omega_{\nu, n}^2$ values changes with $D$. In particular, we present results for $D=0.5$ (orange curve), $D=0.2$ (red curve), $D=0.1$ (green curve) and $D=0.06$ (purple curve). For $D=0.5$ the distribution has a chapeau-like shape with somewhat higher values at the edges of the plateau ($\omega_{\nu, n}^2 \approx 0.2$ and $\approx 0.8$). As $D$ decreases, leading system \eqref{eq:LDKG} to a less disordered form, the distribution develops a `bowl' shape feature at its central part, which deepens for smaller values of $D$, while at the same time the peaks at the distribution's edges become higher and their separation distance grows.

After discussing the characteristics of frequencies $\omega_{\nu, n}^2$ \eqref{eq:nomr_freq}, let us focus on the NMs' spatial features. Several approaches can be used to numerically estimate the extent of NMs \cite{Kramer1993,Krimer2010}.
Here, following \cite{Krimer2010}, we consider two main quantities for that purpose: the NMs' localization volume $V_{\nu}$ and their participation number $P_{\nu}$. In particular, we estimate the effective distance between the exponential tails of NMs as
\begin{equation}
\label{eq:V}
	V_{\nu} = \sqrt{12m_2 ^{(\nu)}} + 1,
\end{equation}
where $m_2^{(\nu)}=\sum_l(l_0-l)^2|A_{\nu,l}|^2$
is the NM's second moment. The NM's participation number, which measures the number of highly excited sites in the mode, is given by
\begin{equation}
\label{eq:P}
	P_{\nu} = \frac{1}{\sum_l\lvert A_{\nu, l}\rvert ^4}.
\end{equation}
These two quantities were found to correctly capture the main features of the NM's extent as they are proportional to the average localization length \eqref{eq:loc_length_LDKG}, which can be computed through the transfer-matrix approach  \cite{Kramer1993,Krimer2010}. Since here we want to focus on the statistical properties of the NMs' spatial features, we compute   $V_{\nu}$ \eqref{eq:V} and $P_{\nu}$ \eqref{eq:P} for various values of $D$.

\begin{figure}
	\includegraphics[width=0.5\textwidth,keepaspectratio]{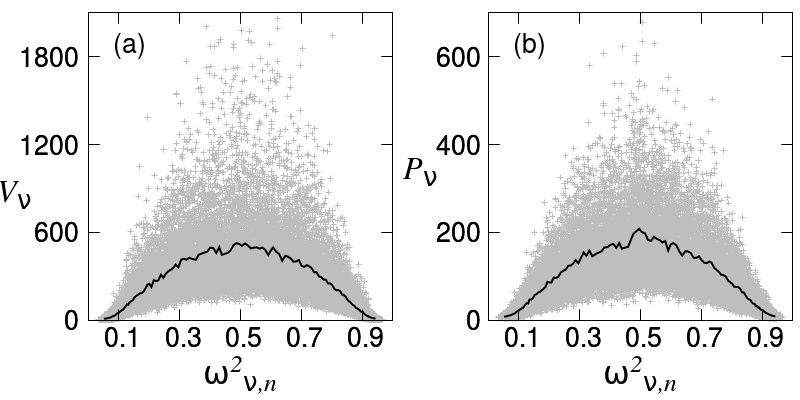}
	\caption{\label{fig:PV_one}
(a) The localization volume $V_{\nu}$ \eqref{eq:V} and (b) the participation number $P_{\nu}$ \eqref{eq:P} of the NMs of system \eqref{eq:LDKG} with $D=0.1$ for $n_d=100$ disorder realizations, with respect to the NMs' normalized
		squared frequency $\omega^2_{\nu,n}$ \eqref{eq:nomr_freq}. In both panels only NMs whose mean position $l_0$ is in the central one-third of the lattice extent are considered. The black curves correspond to the running averages of the plotted quantities, i.e.~(a) $\langle V\rangle$, and (b) $\langle P\rangle$.
}
\end{figure}

\begin{figure}
	\includegraphics[width=0.4\textwidth,keepaspectratio]{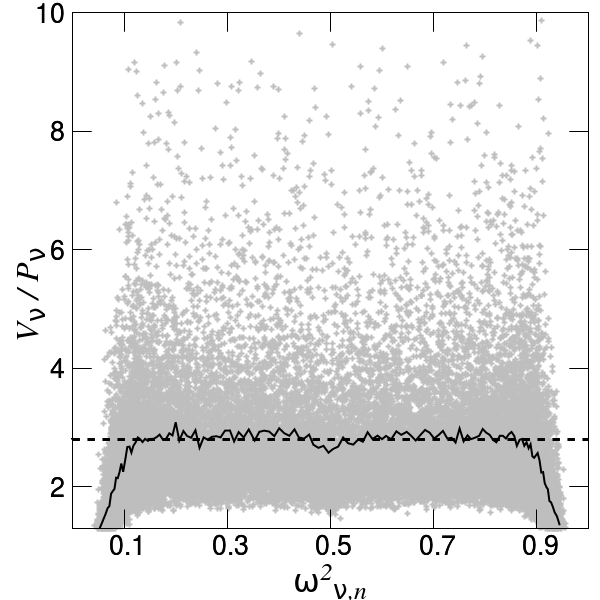}
	\caption{\label{fig:PV_ratio} The ratio $V_{\nu}/P_{\nu}$ of the results of   Fig.~\ref{fig:PV_one} as a function of the normalized squared frequency $\omega^2_{\nu,n}$ of the NMs. The black curve corresponds to the running average of the presented data, while the horizontal dashed line indicates the value $V_{\nu}/P_{\nu}=2.8$.
}
\end{figure}

In Fig.~\ref{fig:PV_one} we plot the localization volume $V_{\nu}$ \eqref{eq:V} [Fig.~\ref{fig:PV_one}(a)] and the participation number $P_{\nu}$ \eqref{eq:P} [Fig.~\ref{fig:PV_one}(b)] of the NMs of $n_d=100$ disorder realizations of Hamiltonian \eqref{eq:LDKG} with $D=0.1$, as a function of the normalized squared frequency $\omega^2_{\nu,n}$ \eqref{eq:nomr_freq}. In order to avoid boundary effects we consider only modes whose mean position $l_0$ is in the central one-third of the lattice. The black continuous curves  correspond to running averages  $\langle V\rangle$ [Fig.~\ref{fig:PV_one}(a)] and  $\langle P\rangle$ [Fig.~\ref{fig:PV_one}(b)] of, respectively, quantities $V_{\nu}$ and $P_{\nu}$. In accordance to \cite{Krimer2010} and as expected from the behavior of the localization length  [see Eqs.~\eqref{eq:loc_length_LDKG} and \eqref{eq:loc_length_LDKG_max}] both $\langle V\rangle$  and  $\langle P\rangle$ obtain their maximum values at the center of the frequency band, i.e.~for $\omega^2_{\nu,n}\approx 0.5$. The existence of a scaling relation between $\langle V\rangle$  and  $\langle P\rangle$ (and between  $V_{\nu}$ and  $P_{\nu}$), especially for the middle part of the frequency band where the most extended NMs are, is apparent in  Fig.~\ref{fig:PV_one}, but becomes more evident in Fig.~\ref{fig:PV_ratio} where we plot the ratio $V_{\nu}/P_{\nu}$ versus $\omega^2_{\nu,n}$ for the results of Fig.~\ref{fig:PV_one}. The running average of the data presented in  Fig.~\ref{fig:PV_ratio} (black curve) indicates that, apart from the NMs at the edges of the frequency band, the ratio   $V_{\nu}/P_{\nu}$ is close to $V_{\nu}/P_{\nu}\approx 2.8$.

\begin{figure}
	\includegraphics[width=0.5\textwidth,keepaspectratio]{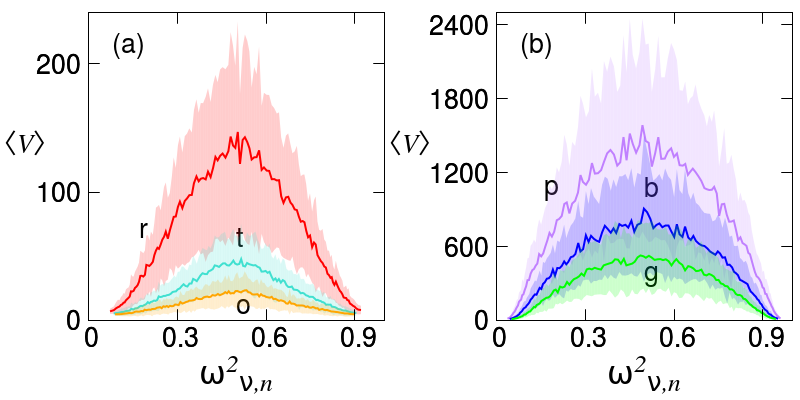}
	\caption{\label{fig:V_many} The average localization volume $\langle V\rangle$ of NMs located at the central one-third of the lattice as a function of  the normalized squared frequency $\omega^2_{\nu,n}$ for (a) $D=0.5$ (orange curve - `o'), $D=0.35$ (turquoise curve - `t') and $D=0.2$ (red curve - `r'), and (b) $D=0.1$ (green curve - `g'), $D=0.08$ (blue curve - `b')  and $D=0.06$ (purple curve - `p'). The shaded area around each curve indicates 1 standard deviation.
}
\end{figure}

After quantifying the spatial extent of the NMs through $V_{\nu}$ (or equivalently through $P_{\nu}$) we investigate in Fig.~\ref{fig:V_many} the effect of $D$ on these results. In particular, we present there how the plot of $\langle V\rangle$ versus $\omega^2_{\nu,n}$ changes with respect to $D$ considering the cases $D=0.5$ (orange curve), $D=0.35$ (turquoise curve) and $D=0.2$ (red curve) in Fig.~\ref{fig:V_many}(a) and $D=0.1$ (green curve), $D=0.08$ (blue curve)  and $D=0.06$ (purple curve) in Fig.~\ref{fig:V_many}(b). For each case the shaded area around the $\langle V\rangle$ curve indicates one standard deviation. In obtaining these results we consider (as in Figs.~\ref{fig:PV_one} and \ref{fig:PV_ratio}) NMs with mean position in the middle one-third of the lattice. From the results of Fig.~\ref{fig:V_many} we observe that both the average value $\langle V\rangle$ and the corresponding standard deviation increase as $D$ decreases.

\begin{figure}
	\includegraphics[width=0.45\textwidth,keepaspectratio]{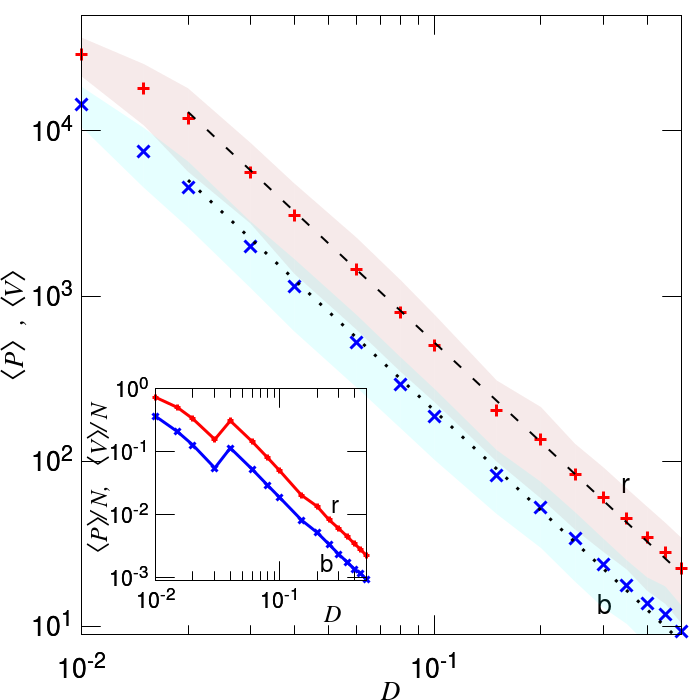}
	\caption{\label{fig:PV_D} The average, over $n_d=100$ disorder realizations, localization volume $\langle V\rangle$ (red points - `r') and participation number $\langle P\rangle$ (blue points - `b') of the NMs located at the central one-third of the lattice and in the middle one-third of the frequency band, as a function of $D$. The shaded area  indicates 1 standard deviation. The straight lines correspond to the functions $\langle V\rangle= a_v /D^2$ (dashed upper curve) and  $\langle P\rangle= a_p /D^2$ (dotted lower curve) with $a_v=5.21$ and $a_p=2.01$. Inset: the ratio $ \langle V\rangle /N$ (upper curve) and $ \langle P\rangle /N$ (lower curve) versus $D$, where $N$ is the lattice size of system \eqref{eq:LDKG} used for the computation of the NMs. Both plots have logarithmic axes.
}
\end{figure}

In order to better quantify the relation between the disorder parameter $D$ and the average spatial extent of the NMs, we restrict our analysis to the more extended modes of the system by considering only NMs in the middle one-third of the frequency band, which at the same time (as we did so far) are located at the central one-third of the lattice. For these modes we compute the average localization volume $\langle V\rangle$ and participation number $\langle P\rangle$ (along with an estimation of the uncertainties of these quantities quantified by their standard deviation) for $n_d=100$ disorder realizations and present them in Fig.~\ref{fig:PV_D}. There we clearly see the increase of $\langle V\rangle$ and $\langle P\rangle$ when $D$ decreases, i.e.~as system \eqref{eq:LDKG} approaches an ordered lattice.

From  the proportionality of $\langle V\rangle$ and $\langle P\rangle$ to the NMs' localization length \cite{Krimer2010}, as well as  from Eqs.~\eqref{eq:loc_length_LDKG} and \eqref{eq:loc_length_LDKG_max}, we expect for both quantities to scale as $\propto D^{-2}$. This is indeed true as the data of $\langle V\rangle$ ($\langle P\rangle$) in Fig.~\ref{fig:PV_D} are well fitted by the function $a_v/D^2$ ($a_p/D^2$) with $a_v=5.21 \pm 0.09$ ($a_p=2.01 \pm 0.05$) shown by a dashed (dotted) straight line in Fig.~\ref{fig:PV_D}. This fitting sets the ratio $\langle V\rangle/\langle P\rangle \approx 2.6$, which is quite close to the value $2.8$ obtained in Fig.~\ref{fig:PV_ratio} for one particular $D$ value. We note that although in Fig.~\ref{fig:PV_D} we present results for $0.01 \leq D \leq 0.5$ we use only the points with $0.02 \leq D \leq 0.5$ for the fittings. We do so because the results with very small $D$ values (namely $D=0.01$ and $D=0.015$) presented in Fig.~\ref{fig:PV_D} should be considered with caution as the extent of the NMs for these cases is quite large with respect to the considered lattice size $N=50{,}000$ and consequently the NMs' properties, as well as the $ \langle V\rangle$ and $ \langle P\rangle$ values, might be influenced by the presence of the boundaries. In order to substantiate this argument we present in  the inset of Fig.~\ref{fig:PV_D}  the values of $ \langle V\rangle /N$ and $ \langle P\rangle /N$ for the performed simulations. Based on these results we decided to perform the fittings in Fig.~\ref{fig:PV_D} considering only data for which $ \langle P\rangle /N \lesssim 10^{-1}$ in order to exclude potential influences of  the lattice boundaries to the NMs' shapes and properties. Let us note here that, in order to obtain $ \langle P\rangle /N \approx 10^{-1}$ for the $D=0.015$ and/or $D=0.01$ cases we would need to find the eigenvalues and the eigenvectors of matrices ${\bf A}$ \eqref{eq:A_epsilon} with dimensions of the order of $10^6 \times 10^6$, which is an extremely hard computational task.

\section{Summary and discussion}
\label{sec:summary}

In this work we studied the properties of the NMs of a modified 1D disordered Klein-Gordon chain, whose disorder strength can be adjusted through two parameters. The first, $D$, is directly influencing disorder as it defines the range of the interval from which the random coefficients of the on-site potential are chosen, while the second, $W$, defines the importance of the nearest-neighbor interactions and in this way it is indirectly influencing the disorder strength.

In our study we fixed $W=4$ and let $D$ approach zero in order to investigate the dynamical changes of the NMs when system \eqref{eq:LDKG} becomes less disordered. We observed that as $D$ decreases the NMs become more extended and their squared frequencies tend to cluster at the edges of the frequency band as their distribution develops a `U'-like shape. We also computed numerically the localization volume $V_{\nu}$ \eqref{eq:V} and the participation number $P_{\nu}$ \eqref{eq:P} of the
modes for different values of $D$ and obtained for their average values, respectively $\langle V\rangle$ and $\langle P\rangle$, the laws $\langle V\rangle \propto  D^{-2}$ and $\langle P\rangle \propto  D^{-2}$ with a scaling $\langle V\rangle \approx 2.6\langle P\rangle$ for $D\leq 0.5$.

The DKG Hamiltonian, i.e.~the nonlinear version of Hamiltonian \eqref{eq:LDKG}, is a physically relevant system as it can, for example, model atomic arrays subject
to external fields, e.g.~anharmonic lattice vibrations in
crystals \cite{Ovchinnikov2001}. Thus, the introduction of the $D$ parameter to control its disorder, alongside $W$, provides us with more flexibility in the way we can tune the model. In particular, it allows us to influence the system's disorder strength either by changing the linear part of the on-site potential (different $D$ values) and/or the power of the nearest-neighbor interactions (different $W$ values). Thus, this separation could allow us  to investigate the effect of different physical processes on the system's dynamics, in ways which could also be realized experimentally.

The two scales which will determine the wave packets' evolution in the presence of nonlinearity in the DKG model are the width $\Delta_K$ \eqref{eq:Delta_K} of the LDKG system's spectrum and the average spacing $d$ of the squared frequencies inside the NMs' localization volume. For $W=4$ these quantities become  
\begin{equation}
\label{eq:Delta_K_W4}
	\Delta_K = 2D + 1,
\end{equation}
and 
\begin{equation}
\label{eq:d_W4}
	d=\frac{\Delta_K}{\langle V\rangle} \approx \frac{D^2 (2D+1)}{5.21},
\end{equation}
where the fitting $\langle V\rangle= 5.21/D^2$ shown in Fig.~\ref{fig:PV_D} was used in obtaining \eqref{eq:d_W4}. Thus, our analysis constitutes the first step toward understanding in more depth the influence of disorder on the chaotic behavior of the DKG system when $D$ is changed, something which we plan to address in a future publication.

\acknowledgments
B.~S.~was funded by the Muni
University ADB-HEST staff development fund.
J.-J.~d.~P. was supported by the Standard Bank of South Africa and the University of Cape Town (UCT). B.~M.~M.~was supported by the National Research Foundation of
South Africa. Ch.~S.~acknowledges support by the UCT's Research Committee (URC). The authors also  thank the High Performance Computing facility of the UCT and the Center for High Performance Computing (CHPC) for
providing the computational resources needed for obtaining the numerical results of this work, as well as their
user-support teams for their help on many practical issues.



\end{document}